\documentclass[cp1251,12pt]{jetp}

\begin{document}
{\English

\title{Expansion of the irregular solution in the theory of Stark effect
in hydrogenic-like Rydberg atoms}

\affiliation{Institute of Problems of Chemical Physics, Russian Academy of
Sciences \\ 142432 Chernogolovka, Russia}

\author{V.~G.}{Ushakov}
\email{uvg@icp.ac.ru}
\author{V.~I.}{Osherov}
\email{osherovv@icp.ac.ru}
\author{E.~S.}{Medvedev}
\email{medvedev@icp.ac.ru}

\rtitle{Expansion of the irregular solution \dots } %
\rauthor{V.~G.~Ushakov, V.~I.~Osherov, E.~S.~Medvedev }

\abstract{We derive the expansion of the irregular physical solution over the
spherical solutions at negative energies, which is necessary for obtaining the
$S$ matrix of the process. The relation of this expansion to the theory
developed by Giannakees et al., Phys. Rev. A 94, 013419 (2016), is analyzed. In
particular, we show that the expansion of the irregular solution missing in
Giannakees et al.'s theory can be derived from one of their main postulates.
The expansion thus obtained turns out to be numerically equivalent to our
expansion up to high angular momenta. Analytical expressions for the key matrix
of both expansions are derived.}

\maketitle


\section{Introduction}

In order to quantitatively describe the Stark photoabsorption spectra of
alkali-metal atoms, Harmin \cite{Harmin1981} used the local frame
transformation (LFT) approach by Fano \cite{Fano1981}. In frame of this
approach, the "physical" solutions in parabolic coordinates, i.e. the ones that
are bounded at infinity, are to be matched near the core with the spherical
solutions that satisfy the boundary condition defined in the quantum-defect
theory \cite{Seaton1966,Seaton1983}. According to this theory, the solutions
outside the core have the form of a specific linear combination of the regular
and irregular spherical Coulomb functions. In the presence of the electric
field, this remains valid up to the distances $r$ where mixing of the states
with different angular momenta $l$ by the field can be neglected. At larger
distances, the parabolic Stark solutions are to be used. Therefore, in order to
obtain the solution valid within the full range of $r$ outside the core, one
has to match the physical parabolic solution to the spherical solutions with
definite values of $l$. Such matching is possible only locally, within the
range of intermediate distances, where both the effect of core structure and
the influence of the external field are negligible in comparison with the
Coulomb attraction to the core. The matching procedure is based on the mutual
expansions between the regular and irregular parabolic and spherical solutions.
These expansions form the base for calculation of the observable quantities.
Harmin's theory was successfully applied to calculations of the photoionization
cross section of sodium atoms and to interpret the photoionization experiments.
However, its application to calculations of the differential cross section in
experiments on ionization microscopy and comparison with the results of highly
accurate experiments (see e.g. \cite{StevensIuBergemanMetcalf1996}) turned out
to be unsatisfactory. Moreover, it was found that the LFT of the irregular wave
function defined in Harmin's theory does not obey some necessary requirements
\cite{StevensIuBergemanMetcalf1996,ZhaoFabrikantDuBordas2012,
GiannakeasGreeneRobicheaux2015,GiannakeasGreeneRobicheaux2016}.

Giannakees et al. have recently developed a generalized LFT (GLFT)
\cite{GiannakeasGreeneRobicheaux2016} that employs the formal use of a
single-particle potential and the operator algebra. The GLFT approach avoids
the explicit use of the LFT of the irregular solution, which is nevertheless
implicitly present in this theory and can be derived from one of its key
postulates. In this paper, we first present the LFT of the irregular solution
obtained by our approach and derive an explicit analytical expression for the
LFT key matrix (Sec. \ref{Sec2}). Second, we derive the LFT-transformation
matrix for the irregular solution from Giannakees et al.'s GLFT theory and
compare it with our respective matrix (Sec. \ref{Sec3}). We found that the two
matrices give numerically equivalent expansions of the irregular solution.

\section{Expansion of the irregular solution \label{Sec2}}

In the limit of small external field, $F\ll 1$ (atomic units are used), one can
specify the core range of distances, $r \lesssim 1$, and an intermediate,
Coulomb range, $1\ll r\ll F^{-1/2}$, where the potential is Coulombic, $-1/r$,
and the external field is weak, $Fr\ll 1/r$. Yet, mixing of the states with
different angular momenta $l$ by the external field can take place within the
Coulomb region at distances where the field energy is larger than the
centrifugal energy difference between neighboring states $l$ and $l-1$, $Fr\geq
2l/r^2$. Hence, the mixing can be neglected only within the "near-Coulomb"
region next to the core, $1\ll r\ll F^{-1/3}$, which is much narrower than the
Coulomb one. At such small $r$, i.e. in the near-Coulomb region, the spherical
and parabolic solutions coexist so that the LFT between them can be performed,
whereas only the latter exists at large $r$, i.e. outside it. Because of the
$l$-mixing within the Coulomb region, the matched spherical functions must
unavoidably involve a linear combination of states with different momenta. The
resulting matching equation is
\begin{equation}
P_{l}^{m}\left( \cos \theta \right) G_{l}(r)+\sum_{l^{\prime }=m}^{\infty
}\gamma _{l,l^{\prime }}P_{l^{\prime }}^{m}\left( \cos \theta \right)
F_{l^{\prime }}(r)=\sum\limits_{k=1}^{\infty }\Upsilon _{l,k}\psi _{k}\left(
\xi ,\eta \right) ,  \label{IrrMatching}
\end{equation}%
where $P_{l}^{m}\left( \cos \theta \right) $ are the Legendre polynomials, $%
F_{l^{\prime }}(r)$ and $G_{l}(r)$ are the regular and irregular (at $r=0$)
solutions of the radial Schr\"{o}dinger equation in the pure Coulomb potential,
and $\psi _{k}\left( \xi ,\eta \right) $ are the physical, i.e. not increasing
at infinity, irregular parabolic solutions in the pure Stark potential. The
coefficients $\gamma _{l,l^{\prime }}$ and $\Upsilon _{l,k}$
are uniquely defined by the matching conditions. The right-hand side of Eq. (%
\ref{IrrMatching}) represents the irregular physical solution at large $r$,
i.e. everywhere outside the core region. Inside the near-Coulomb region, where
the spherical solutions are simultaneously  exist, it can be locally
transformed to a linear combination of the spherical functions. The solutions
in the parabolic and spherical frames must approximately coincide
locally, i.e. within the near-Coulomb region, as is expressed by Eq. (\ref%
{IrrMatching}). The details of the derivation are given elsewhere \cite%
{UOM2019}.

In Ref. \cite{UOM2019}, matrix $\gamma _{l,l^{\prime }}$ entering Eq. (\ref%
{IrrMatching}) could be calculated only numerically. In this section, we derive
an analytical expression for this matrix. The radial spherical functions in the
left-hand side of Eq. (\ref{IrrMatching}) are defined in Ref. \cite{UOM2019} as
\begin{equation}
F_{l}\left( r\right) =\left( \frac{r}{n}\right) ^{l}e^{-r/n}\Phi \left(
-n+l+1,2l+2,2r/n\right) ,  \label{F_l}
\end{equation}%
and
\begin{equation}
G_{l}\left( r\right) =\left( \frac{r}{n}\right) ^{l}e^{-r/n}\Psi \left(
-n+l+1,2l+2,2r/n\right) ,  \label{G_l}
\end{equation}%
where $n=1 / \sqrt{-2E}$, $E$ is the energy, $\Phi \left( a,b,x\right) $ and $%
\Psi \left( a,b,x\right) $ are Kummer's functions $M\left( a,b,x\right) $ and
$U\left( a,b,x\right) $, respectively \cite{Erdelyi_I_1981}; functions
$F_{l}\left( r \right)$ and $G_{l}\left( r \right)$ are not normalized. We
consider the case of $n \gg 1$, which corresponds to highly excited Rydberg
states.

In the right-hand side of Eq. (\ref{IrrMatching}), functions $\psi _{k}\left(
\xi ,\eta \right) $ are the irregular parabolic solutions of the pure Stark
problem,
\begin{equation}
\psi _{k}\left( \xi ,\eta \right) =\chi _{\nu _{k}}\left( \xi \right) \psi
_{\mu _{k}}\left( \eta \right) .
\end{equation}%
Here, $\chi _{\nu _{k}}\left( \xi \right) $ are the normalized-to-unity
solutions of the eigenvalue problem for the finite motion along the parabolic
coordinate $\xi $ and $\psi _{\mu _{k}}\left( \eta \right) $ are the irregular
parabolic solutions of the Stark equation for the infinite
motion along $\eta $. The quantum numbers $\nu _{k}=n \beta_k-(m+1)/2$ and $%
\mu _{k}=n-\nu _{k}-m-1$ are non-integers. They correspond to the discrete set
of eigenvalues $\beta_k$ of the separation constant $\beta$ (partial charge) of
the separable Schr\"odinger equation in the parabolic coordinates. At small
$r$, i.e. in the near-Coulomb region, these functions approximately coincide
with the Coulomb parabolic solutions,
\begin{equation}
\chi _{\nu _{k}}\left( \xi \right) \approx c_{k}\,f_{\nu _{k}}\left( \xi
\right) ,\quad \psi _{\mu _{k}}\left( \eta \right) \approx g_{\mu _{k}}\left(
\eta \right) ,
\end{equation}%
with $c_{k}$ being the normalization constants. The regular and irregular
Coulomb parabolic functions are defined as%
\begin{equation}
f_{\varkappa }(\zeta )=\left( \frac{\zeta }{n}\right) ^{m/2}e^{-\zeta /2n}\Phi
\left( -\varkappa ,m+1,\zeta /n\right)   \label{freg}
\end{equation}%
where $\zeta =\xi $ and $\varkappa =\nu $ (or, as required below in Eq.
(\ref{RegLFT}),
$\zeta =\eta $ and $\varkappa =\mu $) and%
\begin{equation}
g_{\mu }(\eta )=\left( \frac{\eta }{n}\right) ^{m/2}e^{-\eta /2n}\,\Psi \left(
-\mu ,m+1,\eta /n\right) \label{g_mu}
\end{equation}%
respectively. The irregular spherical and parabolic functions (\ref {G_l}) and
(\ref {g_mu}) are chosen from the condition that the solutions are bounded at
infinity.

Coefficients $\gamma _{l,l^{\prime }}$ and $\Upsilon _{l,k}$ are uniquely
defined by the matching of the physical, i.e. not increasing at infinity
parabolic solution with the spherical solutions in the near-Coulomb region and
by the choice of the functions in the form of Eqs. (\ref{G_l}) and
(\ref{g_mu}). The transformation matrix $\Upsilon _{l,k}$ has the form
\cite{UOM2019}%
\begin{equation}
\Upsilon _{l,k}=\frac{W_{l}}{m!\,N_{lm}}A_{\nu _{k}\mu _{k},l}\,c_{k}\,\Gamma
\left( -\mu _{k}\right)   \label{Upsilon}
\end{equation}%
where%
\begin{equation}
A_{\nu \mu ,l}=\sum_{p=0}^{l-m}\frac{(-1)^{p+m}2^{l}\left( l-m\right) !l!\Gamma
\left( 1+\nu \right) \Gamma \left( 1+\mu \right) \left( m!\right) ^{2}}{\left(
2l\right) !\Gamma \left( 1+\nu -p\right) \Gamma \left( 1+\mu +m-l+p\right)
(l-p)!(l-m-p)!(m+p)!p!}
\end{equation}%
is the LFT matrix for the regular solutions. The regular solutions are given
by the product of two functions $f_{\varkappa }(\zeta )$ defined in Eq. (\ref%
{freg}), and the LFT for them is given by%
\begin{equation}
f_{\nu }\left( \xi \right) \,f_{\mu }\left( \eta \right) =\sum_{l=m}^{\infty
}A_{\nu \mu ,l}P_{l}^{m}\left( \cos \theta \right) F_{l}(r). \label{RegLFT}
\end{equation}
In Eq. (\ref{Upsilon}), $W_{l}$ and $N_{lm}$ are the Wronskian and the
normalization constants for the spherical Coulomb functions,%
\begin{equation}
W_{l}=\frac{n\left( 2l+1\right) !}{2^{2l+1}\Gamma \left( 1+l-n\right) }%
,\quad N_{lm}=\frac{2l+1}{2}\frac{\left( l-m\right) !}{\left( l+m\right) !}.
\label{W_l}
\end{equation}

Our method to derive Eq. (\ref{IrrMatching}) is based on the exact expansion of
the irregular spherical functions over the irregular parabolic Coulomb
solutions,
\begin{equation}
\mathcal{G}_{l}\left( r, \theta \right) \equiv P_{l}^{m}\left( \cos \theta
\right) G_{l}(r)=\sum_{n_{1}=0}^{\infty }B_{l,n_{1}}\,f_{n_{1}}\left( \xi
\right) \,g_{n_{2}}\left( \eta \right) \,, \label{IrrExact}
\end{equation}%
where the transformation matrix $B_{l,n_{1}}$ is given by \cite{UOM2019}%
\begin{equation}
B_{l,n_{1}}=\frac{W_{l}\,}{m!\,N_{lm}}A_{n_{1}n_{2},l}\,N_{n_{1}}^{2}\Gamma
\left( -n_{2}\right) \label{Bmatrix}
\end{equation}%
and $n_{2}=n-n_{1}-m-1$. Parameter $N_{n_{1}}$ is the normalization constant
for the regular parabolic function $f_{n_{1}}\left( \xi \right) $,%
\begin{equation}
N_{n_{1}}=\frac{1}{m!}\sqrt{\frac{\left( m+n_{1}\right) !}{n_{1}!n}}.
\end{equation}

Expansion (\ref{IrrExact}) was derived in Ref. \cite{UOM2019} for arbitrary
integer and non-integer $n$ and for a special, unique choice of the radial
spherical function $G_l \left( r \right)$, Eq. (\ref{G_l}), such that it
exponentially decreased at infinity. Owing to this choice, the spherical
function on the left of Eq. (\ref{IrrExact}) at any fixed $\eta \neq 0$ can be
expanded over the quantized basis of $f_{n_1}\left( \xi \right)$. The
coefficients of this expansion are proportional to the bounded at infinity
parabolic solution (\ref {g_mu}) with $ \mu = n_2 $. Note that any irregular
radial function other than $ G_l \left( r \right) $ increases exponentially and
the corresponding spherical function cannot be expanded over the parabolic
solutions.

The physical irregular Stark wavefunction, which is given by the sum in the
right-hand side of Eq. (\ref{IrrMatching}) with coefficients $\Upsilon_{l,k}$
determined by Eq. (\ref{Upsilon}), at any finite $r$ converges to a regular
function of $\theta$, while the singular (at $r=0$) part of this function
coincides with the singular part of $\mathcal{G}_{l}\left( r, \theta \right)$
(see Ref. \cite{UOM2019}). Therefore, the difference between these two
functions is a regular function of $r$ and $\theta$, which can be expanded in a
series over the regular spherical functions,
\begin{equation}
\Psi _{l,\mathrm{reg}}\left( \xi ,\eta \right) = \sum\limits_{k=1}^{\infty
}\Upsilon _{l,k}\chi _{\nu _{k}}(\xi )\,\psi _{\mu _{k}}(\eta
)-\mathcal{G}_{l}\left( r, \theta \right)= \sum_{l^{\prime }=m}^{\infty }\gamma
_{l,l^{\prime }}P_{l^{\prime }}^{m}\left( \cos \theta \right) F_{l^{\prime
}}(r) \label{Psi-l-Reg}
\end{equation}%
which leads to Eq. (\ref{IrrMatching}).

In Ref. \cite{UOM2019}, matrix $\gamma _{l,l^{\prime }}$ was found numerically.
Here, we will find its analytical representation. To begin with, we note that
the main difficulty in calculating the matrix elements $\gamma _{l,l^{\prime
}}$ is related to the nonuniform convergence of the sum on the left-hand side
of equation (\ref{Psi-l-Reg}), see \cite{UOM2019}. The term by term projection
of this sum onto the Legendre polynomials is impossible because it leads to a
divergent series. The truncated sum has a singularity at $\eta = 0$ and the
contribution of this singularity to the integral does not disappear when the
truncation limit tends to infinity.

To find matrix $\gamma _{l,l^{\prime }}$, we transform function $\Psi
_{l,\mathrm{reg}}$ using Eqs. (\ref{Upsilon}), (\ref{IrrExact}), and
(\ref{Bmatrix}),
\begin{eqnarray}
&&\Psi _{l,\mathrm{reg}}\left( \xi ,\eta \right) = \sum\limits_{k=1}^{\infty
}\Upsilon _{l,k}\chi _{\nu _{k}}(\xi )\,\psi _{\mu _{k}}(\eta
)-\sum_{n_{1}=0}^{\infty }B_{l,n_{1}}\,f_{n_{1}}(\xi )g_{n_{2}}(\eta ) =
\frac{W_{l}}{N_{lm}m!} \times
\notag \\
&&\left[ \sum\limits_{k=1}^{\infty }A_{\nu _{k}\mu _{k},l}\,c_{k}^{2}\,\Gamma
\left( -\mu _{k}\right) \,f_{\nu _{k}}(\xi )g_{\mu _{k}}(\eta
)-\sum_{n_{1}=0}^{\infty }A_{n_{1}n_{2},l}\,N_{n_{1}}^{2}\,\Gamma \left(
-n_{2}\right) \,\,f_{n_{1}}(\xi )g_{n_{2}}(\eta )\right] . \label{PsiReg}
\end{eqnarray}%
Both sums in the right hand side of Eq. (\ref{PsiReg}) converge non-uniformly
at $ \eta = 0 $, and when the truncation limits are chosen arbitrarily, the
difference between these sums turns out to be a singular function that cannot
be expanded over the spherical harmonics. However, this singularity can be
eliminated if specially selected cutoff functions are introduced into the sums.

At asymptotically large values of $k$ and $n_{1}$, coefficients
$A_{\nu_k \mu_k,l}$ and $A_{n_{1}n_{2},l}$ are smooth functions of indices%
\begin{equation}
A_{\nu \mu ,l}\approx (-1)^{l}\nu ^{l-m}\frac{2^{l}\left( m!\right) ^{2}}{%
\left( l!\right) ^{2}\left( l+m\right) !}.
\end{equation}
Product $\Gamma \left( -\mu \right) \,g_{\mu }(\eta )$ at large negative $%
\mu $ is also a smooth function of $\mu $. Due to a smooth dependence of the
terms in the sums, summations for large values of indices can be replaced by
integrations,%
\begin{eqnarray}
&&\sum_{n_{1}=n_{1,\max }}^{\infty }F_{\mathrm{coul}}\left( n_{1}\right)
\,A_{n_{1}n_{2},l}\,N_{n_{1}}^{2}\,\Gamma \left( -n_{2}\right)
\,\,f_{n_{1}}(\xi )g_{n_{2}}(\eta )=  \notag \\
&&\int\limits_{n_{1,\max }}^{\infty }F_{\mathrm{coul}}\left( n_{1}\right)
\,A_{n_{1}n_{2},l}\,N_{n_{1}}^{2}\,\Gamma \left( -n_{2}\right)
\,\,f_{n_{1}}(\xi )g_{n_{2}}(\eta )\,dn_{1},
\end{eqnarray}%
and%
\begin{eqnarray}
&&\sum\limits_{k=k_{\max }}^{\infty }F_{\mathrm{st}}\left( \nu _{k}\right)
\,A_{\nu _{k}\mu _{k},l}\,c_{k}^{2}\,\Gamma \left( -\mu _{k}\right) \,f_{\nu
_{k}}(\xi )g_{\mu _{k}}(\eta )=  \notag \\
&&\int\limits_{\nu _{k_{\max }}}^{\infty }F_{\mathrm{st}}\left( \nu \right)
\,A_{\nu \mu ,l}\,c_{k}^{2}\,\Gamma \left( -\mu \right) \,f_{\nu }(\xi )g_{\mu
}(\eta )\,\frac{dk}{d\nu }\,d\nu .
\end{eqnarray}%
Here $F_{\mathrm{st}}\left( \nu _{k}\right) $ and $F_{\mathrm{coul}}\left(
n_{1}\right) $ are the cutoff functions for the Stark and Coulomb sums. At
large values of $k$ and $n_{1}$, the derivative $dk\,/\,d\nu $ is calculated
as%
\begin{equation}
\frac{dk}{d\nu }=\frac{N_{n_{1}}^{2}}{c_{k}^{2}},
\end{equation}%
where $n_{1}=\nu $. Then, putting $n_{1,\max }=\nu _{k_{\max }}$ and $F_{%
\mathrm{coul}}\left( z\right) =F_{\mathrm{st}}\left( z\right) =F_{\mathrm{cut%
}}\left( z\right) $ makes the above two integrals equal to each other.

The singular behavior of functions $g_{\mu _{k}}$ and $g_{n_{2}}$ in Eq. (%
\ref{PsiReg}) is determined by the confluent hypergeometric function $\Psi
\left( a,b,x\right) $, which can be presented as a sum of a uniformly converged
series in powers of $\eta $ and a finite number of singular terms.
After introducing the universal cutoff function into the sums of Eq. (\ref%
{PsiReg}), the singular terms will cancel and the resulting function can be
expanded over the Legendre polynomials, the expansion of the regular parts of
these sums being carried out in the same way as had been done for the regular
functions in Ref. \cite{UOM2019}. Finally, we get the analytical
expression for $\gamma _{l,l^{\prime }}$,%
\begin{equation}
\gamma _{l,l^{\prime }}=\frac{W_{l}}{N_{lm}}\left[ \sum_{n_{1}=0}^{\infty
}F_{\mathrm{cut}}(n_{1})\,A_{n_{1}n_{2},l}\,N_{n_{1}}^{2}\breve{A}%
_{n_{1}n_{2},l^{\prime }}-\sum\limits_{k=1}^{\infty }F_{\mathrm{cut}}(\nu
_{k})\,A_{\nu _{k}\mu _{k},l}\,c_{k}^{2}\,\breve{A}_{\nu _{k}\mu _{k},l^{\prime
}}\,\right]   \label{GAMMA1}
\end{equation}%
where
\begin{equation}
\breve{A}_{\nu \mu ,l}=\sum_{p=0}^{l-m}\frac{(-1)^{p+m}\,2^{l}\left( l-m\right)
!l!\Gamma \left( 1+\nu \right) \Gamma \left( 1+\mu +m\right) \,\Psi \left( \mu
,l-m-p\right) }{\left( 2l\right) !\Gamma \left( 1+\nu -p\right) \Gamma \left(
1+\mu +m-l+p\right) (l-p)!(l-m-p)!(m+p)!p!} \label{At}
\end{equation}%
and
\begin{equation}
\Psi \left( \mu ,s\right) =\psi \left( -\mu +s\right) -\psi \left( 1+m+s\right)
-\psi \left( 1+s\right)
\end{equation}%
($\psi $ is the digamma function).

Note that convergence of the sums in Eq. (\ref{GAMMA1}) is provided by the
cutoff functions, which have the same functional form in both sums. Note also
that the true matching of the Stark wave function to the spherical Coulomb
solutions is realized only in the asymptotic limit of $n\rightarrow \infty $.
At large yet finite $n$, the exact cancelation of residuals of sums in Eq.
(\ref{PsiReg}) is impossible. This circumstance reflects the approximate nature
of the matching, which fully neglects the external field in the near-Coulomb
region. The accuracy of matching is also influenced by an important physical
parameter
\begin{equation}
\delta =16Fn^{4}, \label{delta}
\end{equation}%
which determines the height of the potential barrier for ionization: $\delta
=1$ corresponds to the classical ionization threshold. In practice, when $%
\delta $ is on the order of unity, a high accuracy of the matching can be
achieved already for $n>10$ by a reasonable choice of the cutoff function.

The new features of our LFT expansion, Eq. (\ref{IrrMatching}), as compared to
Harmin's LFT expansion \cite{Harmin1981} are that the matching of the spherical
and parabolic solutions is performed within the near-Coulomb region, which is
narrower than the full Coulomb one, and that the sum over the regular spherical
solutions in the Coulomb potential is added to the left-hand side of the
matching equation (\ref{IrrMatching}). Outside the near-Coulomb region, strong
$l$-mixing takes place, therefore the physical solution extended to the region
of spherical symmetry must involve the states with various $l$. This
drastically differs from Harmin's matching equation where the physical solution
is matched to a spherical function with a definite $l$ within the full Coulomb
region.

The advantage of our expansion over the respective Harmin's expansion was
demonstrated numerically for $l=1,3,5$, $r=10-80$, and $-1<\cos \theta <1$ by
comparison with the exact solution \cite{UOM2019}.

Using the expansion of the irregular solution (\ref{IrrMatching}) and a
similar expansion of the regular solution \cite%
{Harmin1981,GiannakeasGreeneRobicheaux2016,UOM2019}, we derived the $S$ matrix
\cite{UOM2019} suitable for calculating the observed quantities such as
photoionization cross-sections. A different GLFT approach developed by
Giannakees et al. \cite{GiannakeasGreeneRobicheaux2016} avoids using the
irregular solution. Nevertheless, the explicit expansion of the irregular
solution can be derived from the basic relations of the GLFT theory, which can
be compared with our result. This is performed in the next section.

\section{Comparison with the GLFT theory \label{Sec3}}

In this section, notations of Ref. \cite{GiannakeasGreeneRobicheaux2016} are
used, with prefix "G" added to the equation numbers. One of the basic strong
statements of the GLFT theory is the equivalence of two Green functions at
small distances (see Eqs. G18 - G21 and
subsequent text in Ref.  \cite{GiannakeasGreeneRobicheaux2016}),%
\begin{equation}
G^{C-S,\mathrm{smooth}}\left( \mathbf{r},\mathbf{r}^{\prime }\right) =G^{C,%
\mathrm{smooth}}\left( \mathbf{r},\mathbf{r}^{\prime }\right) .  \label{sic}
\end{equation}

Both functions are defined as divergent sums over the separation constant
$\beta $ (the partial charge) running an infinite set of discrete values, but
the actual summations are performed up to a common maximum value of $\beta $.
Substituting the explicit definitions of the relevant functions given in Eqs.
G13, G15, G18, G19, and G22 and expanding all the regular parabolic functions
over the basis of the spherical harmonics (Eq. G10), we obtain the expansion of
the irregular function in the form%
\begin{equation}
g_{\epsilon \, l m}\left(\mathbf{r}\right)=\sum_{\beta ^{F} } \left[
U^{T}\left( \epsilon \right) \right]_{l \beta ^{F} m}\,\chi _{\epsilon \beta
^{F} m }\left(\mathbf{r}\right)-\sum_{l^{\prime }}\tilde{\gamma}_{l,l^{\prime
}}f_{\epsilon \, l^{\prime } m }\left(\mathbf{r}\right), \label{irreg}
\end{equation}%
where%
\begin{equation}
\tilde{\gamma}_{l,l^{\prime }}=J_{l,l^{\prime }}+\cot \left( \pi n\right)
\label{gamma}
\end{equation}%
and $J_{l,l^{\prime }}$ is given by Eq. G22. Rewriting Eq. (\ref{irreg}) in
terms of our functions and notations, we finally obtain it in the
form of Eq. (\ref{IrrMatching}) with%
\begin{equation}
\gamma _{l,l^{\prime }}=\frac{W_{l}}{N_{lm}\left( m!\right) ^{2}}\left[
\sum_{n_{1}=0}^{\infty }\,A_{n_{1}n_{2},l}\,N_{n_{1}}^{2}\Omega
(n_{2})\,A_{n_{1}n_{2},l^{\prime }}-\sum\limits_{k=1}^{\infty }\,A_{\nu _{k}\mu
_{k},l}\,c_{k}^{2}\,\Omega (\mu _{k})\,A_{\nu _{k}\mu _{k},l^{\prime
}}\,\right] ,  \label{GAMMA}
\end{equation}%
where%
\begin{equation}
\Omega (\mu )=\frac{\Gamma \left( 1+m+\mu \right) }{\Gamma \left( 1+\mu \right)
}\left[ \frac{\psi \left( 1+m+\mu \right) +\psi \left( 1+\mu \right) -2\ln
n}{2}+\pi \cot \left( \pi \mu \right) \right] .  \label{Omega}
\end{equation}%
Both sums in Eq. (\ref{GAMMA}) are divergent, therefore the actual summation is
performed up to a common maximum value of $\beta $ using a suitable cutoff
function.

\begin{figure}[h]
\includegraphics[scale=0.5]{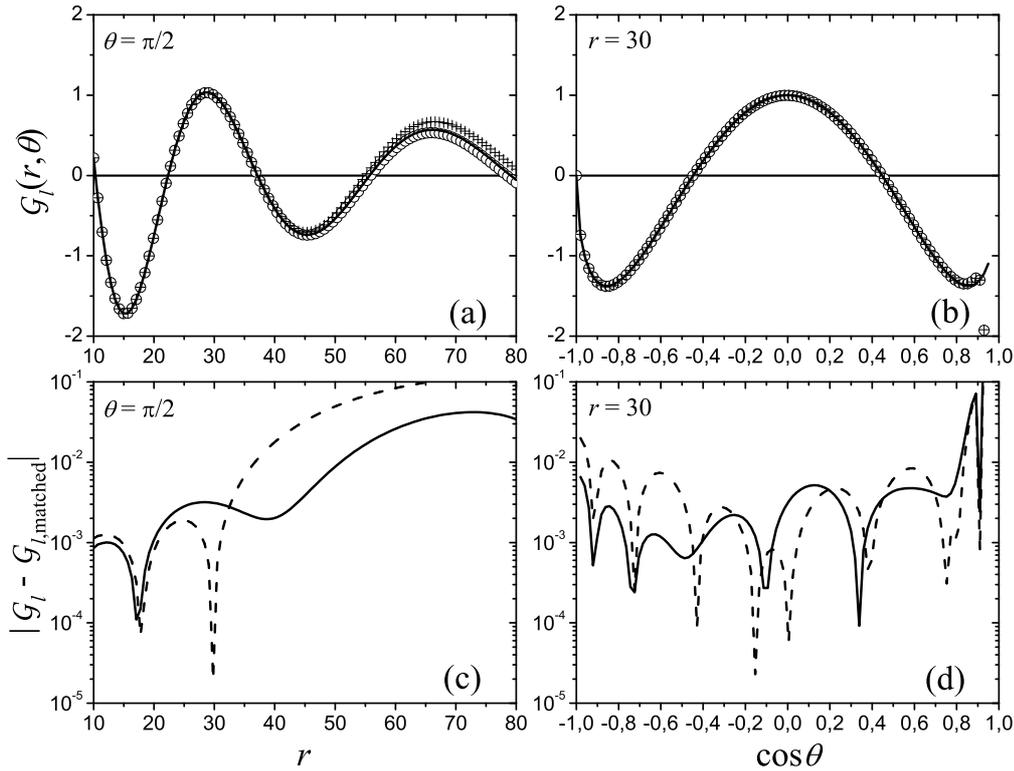}
\caption{Comparison of two matched irregular spherical solutions with the exact
solution at $n=10.5$, $m=1$, $l=3$, and $\delta=1.3$. Full line and symbols at
the upper panels (a) and (b), the exact and matched solutions, respectively.
Since the matched solutions are indistinguishable at this scale, their
differences with the exact solution are shown at the lower panels (c) and (d)
by full and dashed lines for matrices (\ref{GAMMA1}) and (\ref{GAMMA}),
respectively. All functions are divided by the value of $\mathcal{G}_l \left(
r=30, \cos \theta=0 \right)$.} \label{Fig1}
\end{figure}
\begin{figure}[h]
\includegraphics[scale=0.5]{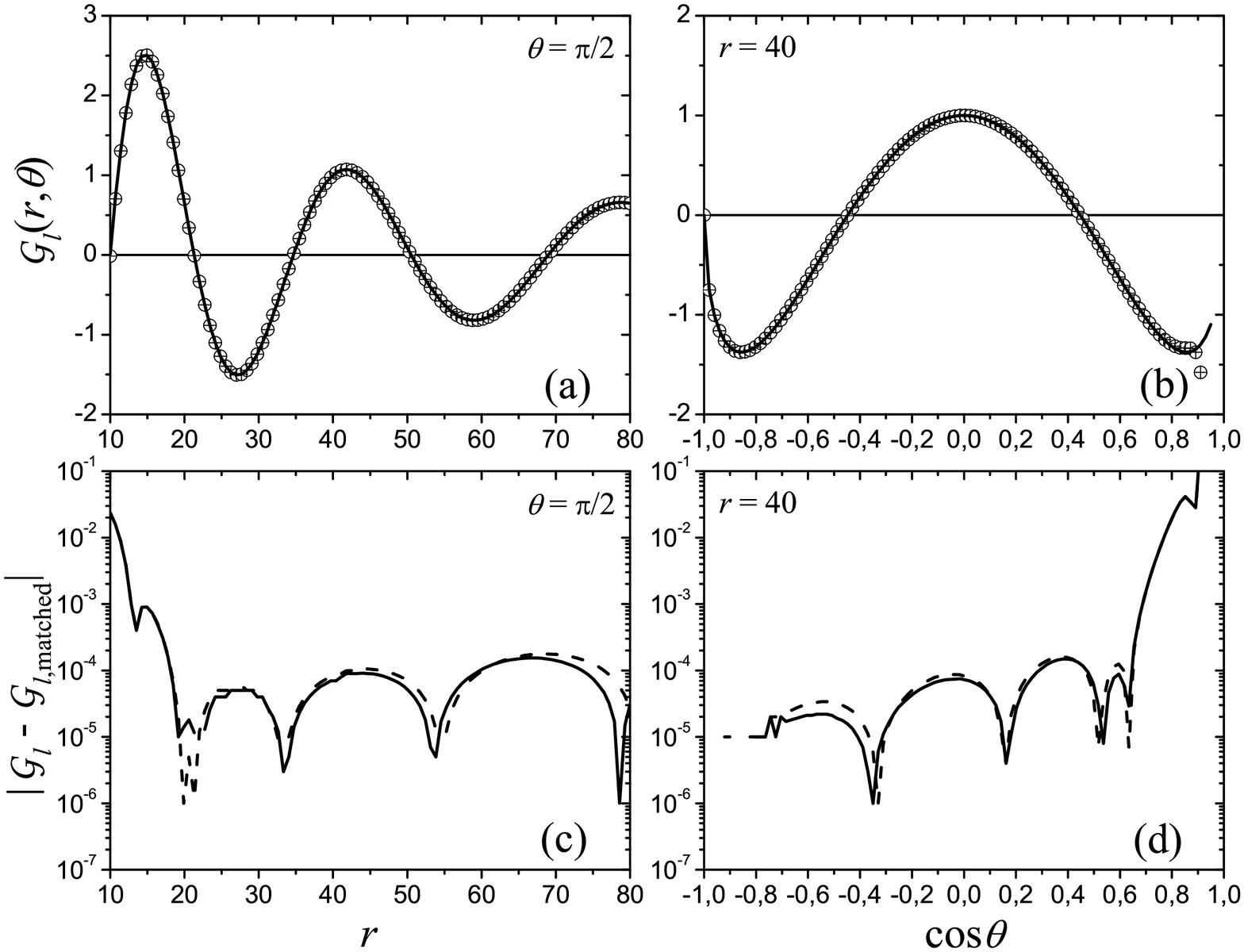}
\caption{The same as in Fig. \ref{Fig1} at $n=28.5$, $m=1$, $l=3$, and
$\delta=1.3$. All functions are divided by $\mathcal{G}_l \left( r=40, \cos
\theta=0 \right)$} \label{Fig2}
\end{figure}

Despite the apparent difference between expressions (\ref{GAMMA1}) and
(\ref{GAMMA}) for $\gamma$, both matrices turn out to be numerically equivalent
up to high angular momenta. In order to compare $\gamma _{l,l^{\prime }}$ from
Eqs. (\ref{GAMMA1}) and (\ref{GAMMA}), we used the cutoff function of Ref.
\cite{GiannakeasGreeneRobicheaux2016}. Two matrices in Eqs. (\ref{GAMMA1}) and
(\ref{GAMMA}) determine two irregular spherical solutions
$\mathcal{G}_{l,\mathrm{matched}}(r,\theta)$ matched to the physical irregular
parabolic solution by two different methods. The upper panels on Figs.
\ref{Fig1} and \ref{Fig2} compare them with the exact function
$\mathcal{G}_{l}(r,\theta)$ whereas the lower ones show the respective
differences. It is seen in Fig. \ref{Fig1} that the matched functions ideally
coincide with the exact function at the selected parameters, their difference
being only on the order of $10^{-3}$ over the most of the variables intervals.

Since the matching is correct only asymptotically at $ n \rightarrow \infty $
(i.e. $F \rightarrow 0$   at fixed $\delta$  of Eq. (\ref{delta})), Fig.
\ref{Fig2} demonstrates the rate of convergence when $n$ increases from $10.5$
to $28.5$. The difference of both matched functions with the exact function
drops down to $10^{-4}$.

Note that the discrepancy between the exact and matched functions in the
figures increases at small $\eta$ ($\cos \theta \approx +1$). This discrepancy
is not a consequence of the matching error as such, rather it is determined by
using a finite basis when calculating the sum in the right-hand side of Eq.
(\ref{IrrMatching}) and by the non-uniform convergence of this sum at $\eta=0$.

\section{Discussion}

The major source of inaccuracies of the Harmin LFT theory is the approximation
that enables matching, at short $r$, of the irregular physical Stark
wavefunction to the irregular spherical solution with a definite value of the
orbital angular momentum $l$. However, such a matching is, strictly speaking,
physically impossible because essential $l$-mixing takes place at short $r$
despite the fact that the external field is weak as compared with the Coulomb
one. Giannakees et al. derived a general expression for the $K$ matrix (i.e.
the real scattering matrix which couples the standing waves at infinity, see
Eq. G6 in Ref. \cite{GiannakeasGreeneRobicheaux2016}) without using the
explicit matching of the irregular solution. They introduced a one-particle
potential modelling the quantum-defect boundary condition near the core and
invoked the Lippmann-Schwinger formalism. In fact, as was demonstrated in the
present paper, the matching of the irregular Stark and Coulomb solutions had
been implicitly present in Giannakees et al.'s theory, and here we have shown
how it can be deduced from one of its basic postulates, Eq. G22, with the
matching being governed by Eqs. (\ref{irreg}) and (\ref{gamma}). Using these
equations together with the expansion of the regular solutions, Eq. G10, one
can obtain exactly the same $K$ matrix as in Ref.
\cite{GiannakeasGreeneRobicheaux2016} without invoking any one-particle
potential. The key matrix $\gamma$ in our notations has the form of Eq.
(\ref{GAMMA}). Our alternative approach is based on the exact expansion of the
irregular spherical Coulomb function over the irregular parabolic Coulomb
solutions, Eq. (\ref{IrrExact}), derived in Ref. \cite{UOM2019}. The resulting
Eq. (\ref{GAMMA1}) for $\gamma$ seemingly differs from Eq. (25). Yet, the
numerical tests showed that the two expressions have similar accuracy and the
identical ranges of validity.

\section{Conclusion}

We performed an analytical matching, in the region of spherical symmetry, of
the physical irregular solution of the Stark problem in the hydrogen-like
Rydberg atoms to a linear combination of the irregular and regular spherical
solutions in the pure Coulomb field. We took into account the fact earlier
ignored by researchers in the field that mixing of the states with different
momenta occurs within the Coulomb region where the external field is weak as
compared to the Coulomb one. With the well-known similar matching for the
regular solution, the $S$ matrix can be constructed using standard procedures
\cite{UOM2019}.

This paper is dedicated to the centenary anniversary of the outstanding
scientist, the Founder and long-term Director of the L. D. Landau Institute for
Theoretical Physics, and merely charming person Academician Isaak Markovich
Khalatnikov, "Khalat" among the colleagues and students. One of us (E. S.
Medvedev) preserves warm memories of 1957-1963 studies at Moscow
Physical-Technical Institute and P. L. Kapitza Insitute of Physical problems
where Khalat was supervisor of the 722 group.

This work was performed in accordance  with  the state task, state registration
No. 0089-2019-0002.


}

\end{document}